\documentclass[aps,pra,twocolumn,showpacs,amsmath,amssymb,superscriptaddress,reprint]{revtex4-1}

\usepackage[utf8]{inputenc}
\usepackage{graphicx}% Include figure files
\usepackage[colorlinks=true,citecolor=blue]{hyperref}

\usepackage{units}

\newcommand{\up}{\uparrow}
\newcommand{\down}{\downarrow}
\newcommand{\kBT}{k_\text{B}T}

\begin{document}
\title{A Topological SQUIPT based on helical edge states in proximity to superconductors}

\author{Lennart Bours}
\affiliation{NEST, Istituto Nanoscienze--CNR and Scuola Normale Superiore, Piazza San Silvestro 12, 56127 Pisa, Italy}
\author{Bj\"orn Sothmann}
\affiliation{Theoretische Physik, Universit\"at Duisburg-Essen and CENIDE, D-47048 Duisburg, Germany}
\author{Matteo Carrega}
\affiliation{NEST, Istituto Nanoscienze--CNR and Scuola Normale Superiore, Piazza San Silvestro 12, 56127 Pisa, Italy}
\author{Elia Strambini}
\affiliation{NEST, Istituto Nanoscienze--CNR and Scuola Normale Superiore, Piazza San Silvestro 12, 56127 Pisa, Italy}
\author{Ewelina M. Hankiewicz}
\affiliation{Institute for Theoretical Physics and Astrophysics, University of W\"urzburg, Am Hubland, D-97074 W\"urzburg, Germany}
\author{Laurens W. Molenkamp}
\affiliation{Experimentelle Physik 3, Physikalisches Institut, University of W\"urzburg, Am Hubland, D-97074 W\"urzburg, Germany}
\author{Francesco Giazotto}
\affiliation{NEST, Istituto Nanoscienze--CNR and Scuola Normale Superiore, Piazza San Silvestro 12, 56127 Pisa, Italy}

\date{\today}
\begin{abstract}
We propose a device based on a topological Josephson junction where the helical edge states of a two-dimensional topological insulator are in close proximity to two superconducting leads. The presence of a magnetic flux through the junction leads to a Doppler shift in the spectrum of Andreev bound states, and affects the quantum interference between proximized edge states. We inspect the emergent features, accessing the density of states through a tunnel-coupled metallic probe, thus realizing a Topological Superconducting Quantum Interference Proximity Transistor (TSQUIPT). We calculate the expected performances of this new device, concluding that it can be used as a sensitive, absolute magnetometer due to the voltage drop across the junction decaying to a constant value as a function of the magnetic flux. Contrary to conventional SQUID and SQUIPT designs, no ring structure is needed. The findings pave the way for novel and sensitive devices based on hybrid devices that exploit helical edge states.
\end{abstract}

\maketitle

\section{Introduction}
The past decade has seen an immense interest in the physics of topological insulators (TIs), motivated by their potential applications in nano-electronics, spintronics and quantum computation~\cite{hasan_colloquium:_2010,qi_topological_2011,hasan_three-dimensional_2011,ando_topological_2013,review_ewelina_2013}.
The existence of both two-dimensional (2D) and three-dimensional (3D) TIs has been predicted roughly ten years ago~\cite{kane_quantum_2005,kane_z2_2005,bernevig_quantum_2006,bernevig_quantum_2006-1,moore_topological_2007,fu_topological_2007,liu_quantum_2008,roy_topological_2009} and was confirmed experimentally shortly afterwards via transport~\cite{konig_quantum_2007} and spectroscopic measurements~\cite{hsieh_topological_2008,xia_observation_2009}.
Two-dimensional TIs have been realized in HgTe/CdTe~\cite{konig_quantum_2007,roth_nonlocal_2009,brune_spin_2012,nowack_imaging_2013,konig_spatially_2013} and InAs/GaSb~\cite{knez_evidence_2011,nichele_insulating_2014,knez_observation_2014,qu_electric_2015,li_observation_2015,du_robust_2015,mueller_nonlocal_2015} quantum wells which exhibit an insulating behavior in the bulk while transport properties are governed by topologically protected gapless edge states. Due to the strong spin-orbit coupling, the spin and momentum degrees of freedom in the edge states are locked, resulting in helical edge channels~\cite{wu_helical_2006}. This means that each surface of a 2D TI hosts one pair of counterpropagating edge states with opposite spin~\cite{brune_spin_2012}. The helical edge states are protected against backscattering by time-reversal symmetry which guarantees robustness against several kinds of disorder and perturbations.
A large number of studies has been put forward on the nature of helical edge states of TIs, including the role of e-e interactions~\cite{schmidt_current_2011,ronetti_spin-thermoelectric_2016,calzona_time-resolved_2016}, breaking of time-reversal symmetry~\cite{delplace_magnetic-field-induced_2012,edge_z_2_2013,vayrynen_helical_2013}, and spin properties~\cite{das_spin-polarized_2011}.

Superconducting correlations can be induced in edge channels via the proximity effect from a conventional superconductor~\cite{maier_induced_2012,williams_unconventional_2012,knez_andreev_2012,oostinga_josephson_2013,finck_phase_2014,hart_induced_2014,pribiag_edge-mode_2015}. The induced superconductivity exhibits both spin-singlet $s$-wave pairing as well as spin-triplet $p$-wave pairing due to the strong spin-orbit coupling in the TI. 
In a topological Josephson junction where two superconducting electrodes are coupled to each other via the edge states of a TI, the presence of $p$-wave pairing opens the possibility to create Majorana modes in the form of topologically protected zero-energy Andreev bound states~\cite{fu_superconducting_2008,tkachov_helical_2013}.
These Majorana modes could give rise to a $4\pi$-periodic Josephson current~\cite{fu_josephson_2009,badiane_nonequilibrium_2011,beenakker_fermion-parity_2013,crepin_parity_2014}, feature which has been observed in experiment recently~\cite{wiedenmann_4-periodic_2016,bocquillon_gapless_2017,deacon_josephson_2017}. Furthermore, they are responsible for an anomalous current-phase relation~\cite{sochnikov_nonsinusoidal_2015,kurter_evidence_2015} and can be identified by their unique phase-dependent thermal conductance~\cite{sothmann_fingerprint_2016}.

Additional control over the gapless Andreev bound states can be obtained by applying a small magnetic flux through a ballistic 2D topological Josephson junction. The magnetic flux induces a local gradient of the superconducting phase, resulting in a finite Cooper-pair (condensate) momentum along the edge, $p_S$. This modifies the amplitude for Andreev reflections in which an incident electron (hole) from the edge state is reflected as a hole (electron) and, thus, modifies the spectrum of the Andreev bound states forming inside the junction as well as the transport properties of the junction~\cite{tkachov_quantum_2015}. In fact, the electrons at the upper and lower edge acquire different momentum shift under Andreev reflection due to opposite sign of a finite Cooper-pair momentum shift, an effect similar to the Doppler shift. This allows for closing of the induced superconducting gap by small magnetic fields of a few mT.

This effect can be exploited, for instance, to manipulate the thermal conductance of the junction. While the thermal conductance is exponentially suppressed in the presence of a superconducting gap, it is twice the thermal conductance quantum when the gap is closed and both edge channels contribute to energy transport. This enables the operation of such a device as a thermal switch~\cite{sothmann_high-efficiency_2017,Martinez-Perez2014}.

\begin{figure}
\centering
	\includegraphics[width=\columnwidth]{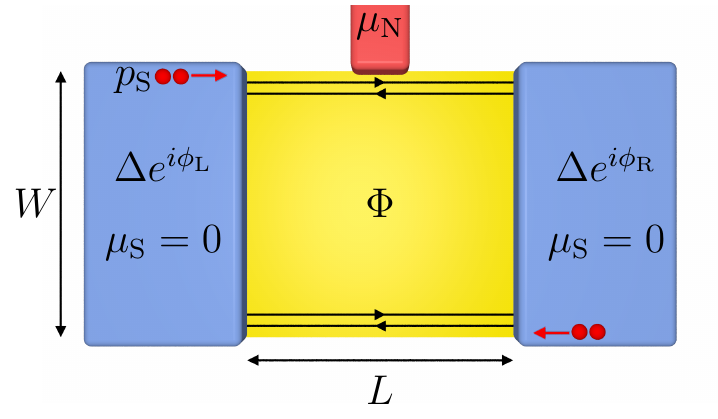}
	\caption{A schematic illustration of the proposed device. A topological Josephson junction is formed by two superconducting leads (blue) coupled via a 2D topological insulator (yellow) of width $W$ and length $L$. Transport through the topological insulator occurs via two helical edge states at the boundary of the insulator. A normal probe terminal (red) is tunnel coupled to the edge states via a tunnel barrier. A magnetic flux $\Phi$ through the junction gives rise to a finite Cooper pair momentum $p_\text{S}$ in the superconducting leads.}
	\label{fig:device}
\end{figure}

In this work, we propose a device to investigate the quantum interference of edge supercurrents in 2D TIs. The quantum interference of supercurrents carried by the two edges of the Josephson junction has been studied by analysing their flux dependence~\cite{baxevanis_even-odd_2015,tkachov_quantum_2015}. Here we aim to probe the density of states of a single proximized edge via a normal tunnel probe, see Fig.~\ref{fig:device}. In essence the proposed device realizes a Topological variant of the Superconducting Quantum Interference Proximity Transistor~\cite{giazotto_superconducting_2010,meschke_tunnel_2011,ronzani_highly_2014,dambrosio_normal_2015,virtanen_spectral_2016,omega_nnano_2016,omega_prb_2017,Giazotto_2011,Ronzani_2017,Ligato_2017,Enrico_2017,Enrico_2016,Jabdaraghi_2017,Jabdaraghi_2014,Jabdaraghi_2016} or TSQUIPT. We characterize the TSQUIPT and its performance, and present calculations of the expected device behavior in transport experiments. 

Not only does the TSQUIPT host physics interesting for fundamental reasons, it can also be used as an absolute magnetometer due to its non periodic flux dependence; by tracing the device response, to an applied flux, one can determine the absolute flux, and by extension the magnetic field to which the device is exposed. Contrary to other flux sensitive devices, no ring structure is needed, which is advantageous with regards to fabrication, and the device sensitivity can be tuned via a bias current.

The paper is organized as follows. In Sec.~\ref{sec:model} we summarize the model and sketch the basic equations for helical edge states in proximity to two superconductors. These are then used to derive the expression for the density of states which will be investigated in detail in Sec.~\ref{sec:characterization} by inspecting the transport properties of a tunnel-coupled metallic probe. The electrical response of the TSQUIPT will be presented, together with the implementation of an absolute magnetometer, with an optimum sensitivity comparable to state of the art commercial SQUIDs. Future prospects are discussed and results are summarized in Sec.~\ref{sec:conclusion}.

\section{Model and basic definitions}
\label{sec:model}
We consider a topological Josephson junction consisting of two superconducting electrodes connected by a 2D TI of length $L$ and width $W$ as depicted in Fig.~\ref{fig:device}.
% The junction is subject to a perpendicular, homogenous magnetic field which gives rise to a magnetic flux $\Phi$. 
We assume that the width of the 2D TI is so large that the overlap between edge channels from different edges can safely be neglected. 
A normal metal probe is weakly tunnel-coupled to the upper edge of the junction. 
Assuming that both superconductors are kept at the same electrochemical potential, $\mu_\text{S}=0$ one can inject a charge current from the probe into the Josephson junction by applying a bias voltage $V=\mu_\text{N}/e$ to the probe terminal.

The left ($\text{L}$) and right ($\text{R}$) superconducting lead are characterized by a superconducting order parameter $\Delta e^{i\phi_\text{L,R}}$.
We assume that the order parameter changes at the superconductor-topological insulator interface on a length scale shorter than the superconducting coherence length $\xi_0=\hbar v_\text{F}/\Delta$, with the Fermi velocity $v_F$, which allows us to model the spatial variation of the order parameter as $\Delta(x) = \Delta [\Theta(-x-L/2)e^{i\phi_\text{L}} + \Theta(x-L/2)e^{-i\phi_\text{R}}]$, where $\Theta(x)$ is the step function. 
We neglect proximity effects inside the junction that would require a self-consistent evaluation of the order parameter. This is a reasonable approximation since transport through the junction proceeds via the two edge channels only~\cite{beenakker_universal_1991}. 
The proximized non-interacting helical edge states~\footnote{Here for sake of simplicity we limit the discussion to the non-interacting case, neglecting possible e-e interactions within edge channels.} at the upper edge are described by the Bogoliubov-de Gennes Hamiltonian
\begin{equation}
	H_\text{BdG}=\left(\begin{array}{cc} h(x) & i\sigma_y \Delta(x) \\ -i\sigma_y \Delta(x)^* & -h^*(x) \end{array}\right).
\end{equation}
In the above equation $\Delta(x)$ is the superconducting pairing potential, while diagonal terms describe the two edge channels and read
\begin{equation}
	h(x)=v_\text{F}\sigma_x\left(-i\hbar\partial_x+\frac{p_\text{S}}{2}\right)+\sigma_0 \mu,
\end{equation}
with $v_{{\rm F}} $ the Fermi velocity, $\sigma_0$ the identity matrix and $\sigma_j$ the Pauli matrices acting on spin space, and the chemical potential $\mu$. We have introduced
\begin{equation}
\label{eq:cp_momentum}
p_\text{S}=\frac{\pi\xi_0\Delta}{v_\text{F}L}\frac{\Phi}{\Phi_0}
\end{equation}
which denotes a finite Cooper-pair (condensate) momentum along the edge. When the normal lead is attached, as in Fig.~1, it can inject right propagating electrons with momentum $p$ and left-propagating electrons with momentum $-p$ both acquiring the same momentum shift $p_S/2$ during Andreev reflection (for the upper edge)  leading to a relative momentum shift for left and right movers similar to the well-known Doppler shift. $\Phi =WLB$ is the flux through the junction, and $\Phi_0$ is the magnetic flux quantum. 

The Cooper pair momentum relevant for transport properties along the edge channels is determined by the width of the 2D TI weak link as long as the superconductor width exceeds that of the TI. As we take $W$ much larger than the spatial extension of the edge states, the Cooper pair momentum is taken to be the boundary value.
The magnetic field is assumed to be sufficiently small such that no backscattering is induced in the helical edge channels~\cite{spin_hall_backscttering_2010}, and superconductivity is not quenched in the leads.
In addition to inducing a finite Cooper pair momentum, the magnetic field also leads to a coordinate dependence of the phase difference $\phi_\text{R}-\phi_\text{L}$. Taking the junction to be smaller than the Josephson penetration depth (which is a good approximation for nanoscale junctions) we obtain for the phase difference across the upper edge $\phi_{u}=\phi_0 + \pi\frac{\Phi}{\Phi_0}$ where $\phi_0$ denotes the phase difference in the absence of a magnetic flux.

The eigenfunctions of the Bogoliubov-de Gennes Hamiltonian in an infinite superconductor with phase $\phi_i$ in Nambu notation are given by
\begin{align}
	\psi_1(x)&=(u_-,u_-,-e^{-i\phi_i}v_-,e^{-i\phi_i}v_-)^T e^{ik_e x},\\
	\psi_2(x)&=(v_-,v_-,-e^{-i\phi_i}u_-,e^{-i\phi_i}u_-)^T e^{ik_h x},\\
	\psi_3(x)&=(u_+,-u_+,e^{-i\phi_i}v_+,e^{-i\phi_i}v_+)^T e^{-ik_e x},\\
	\psi_4(x)&=(v_+,-v_+,e^{-i\phi_i}u_+,e^{-i\phi_i}u_+)^T e^{-ik_h x},
\end{align}
describing right- moving electron-like, left-moving hole-like, left-moving electron-like and right-moving hole-like quasiparticles, respectively. In the above equations we have introduced 
\begin{align}
	u_\pm=\frac{1}{2}\left(1+\frac{\sqrt{E_\pm^2-|\Delta|^2}}{E_\pm}\right),\\
	v_\pm=\frac{1}{2}\left(1-\frac{\sqrt{E_\pm^2-|\Delta|^2}}{E_\pm}\right),
\end{align}
with $E_\pm=E\pm \frac{v_\text{F}p_S}{2}$ and $k_{e,h}$ the wavevector associated to electron and hole-like quasiparticle, respectively. To find the wavefunction in all three regions we consider the wave-matching approach at the S-2D TI interfaces assuming, for simplicity, ideally transparent interfaces.
Let us consider the case of an electron-like quasiparticle impinging on the junction from the left. The wave functions in the three different regions, i.e., left superconductor (S,l), the central (2DTI), and right superconductor (S,r) of the junction can be written as
\begin{align}
	\psi_{{\rm S,l}}(x)&=\psi_1(x)+r_e\psi_3(x)+r_h\psi_2(x),\\
	\psi_\text{2DTI}(x)&=\sum_ia_i\psi_i(x),\\
	\psi_\text{S,r}(x)&=t_e\psi_1(x)+t_h\psi_4(x).
\end{align}
Here $r_e$, $r_h$, $t_e$ and $t_h$ represent the reflection and transmission coefficients for electron-like and hole-like quasiparticles.
Taking into account the continuity of the wave function at the interfaces, we obtain the wave function in the central region, which provides direct access to the transmission probability of quasiparticles through the junction. We remark that while the normal state transmission of a 2D TI equals unity due to Klein tunneling preventing backscattering in the presence of time-reversal symmetry, the transmission in the superconducting state depends on energy, phase difference and magnetic flux in a nontrivial way, due to interference effects~\cite{sothmann_high-efficiency_2017}. These quantum interferences also manifest themselves in the density of states of the junction.
Writing the wave function of the central region as $\psi_\text{2D TI}=(u_\up,u_\down,v_\up,v_\down)$, the corresponding contribution to the density of states of the upper edge channel inside the junction is given by 
\begin{equation}
\rho(E)=\sum_{k,\sigma,\eta = \pm}\left[|u_\sigma|^2\delta(E-E_{k\eta})+|v_\sigma|^2\delta(E+E_{k\eta})\right],
\end{equation}
where we defined $E_{ k\pm}=\sqrt{( v_\text{F}k\pm\mu)^2+\Delta^2}\pm\frac{v_\text{F}p_\text{S}}{2}$. The $+$ and $-$ refer to left- and right-moving quasiparticles, respectively. Due to the counter-propagating nature of helical edge states, the left and right movers are shifted opposite in energy by $\pm \frac{v_\text{F}p_\text{S}}{2}$.

\begin{figure}
\centering
	\includegraphics[width=\columnwidth]{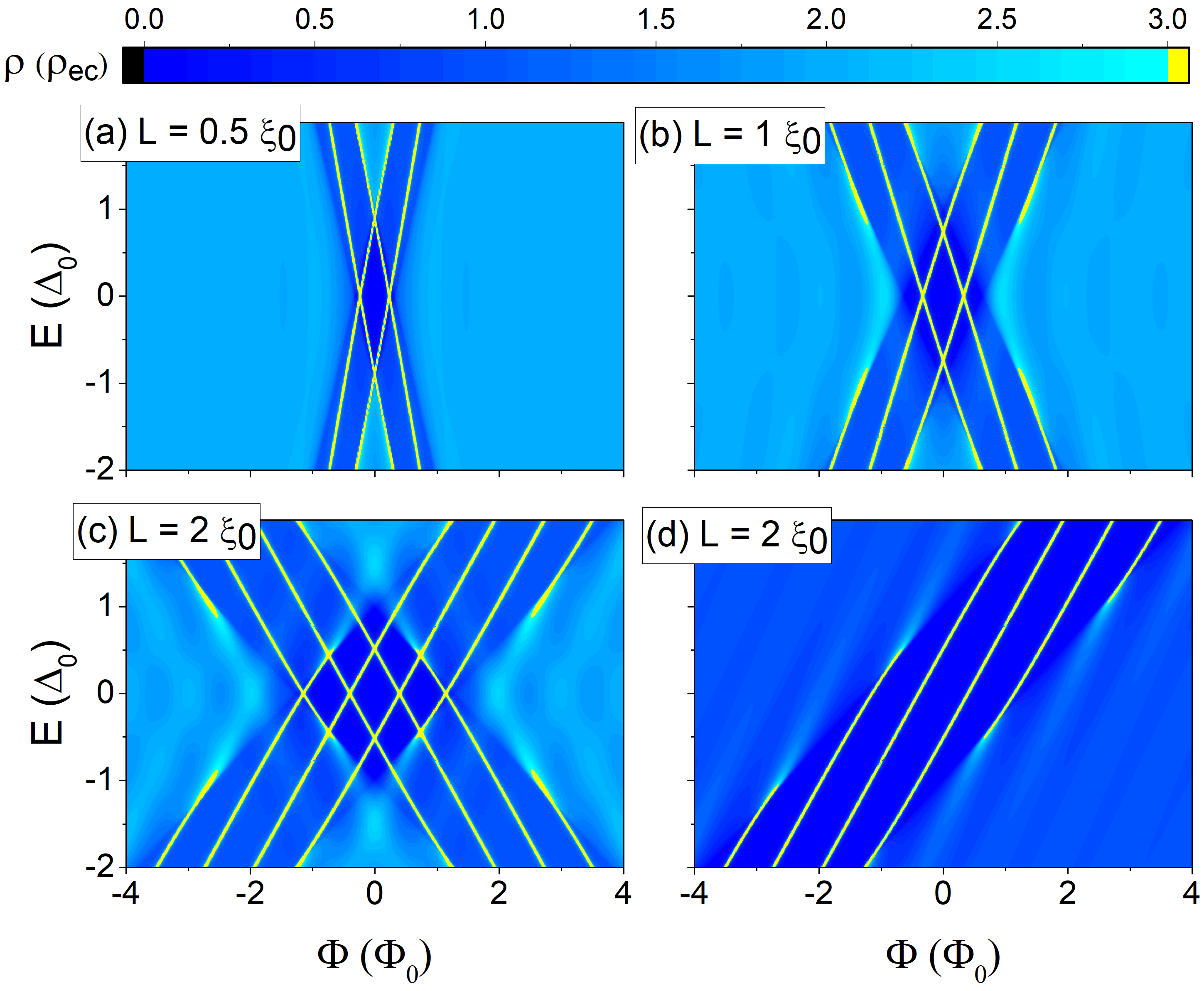}
	\caption{Density of states of the topological Josephson junction as a function of energy, in units of the density of states of a single edge channel in its normal state, and magnetic flux, in units of the flux quantum. Andreev bound states appear as sharp lines inside the gapped regions and transform into broad, decaying resonances outside the gap. The superconducting phase difference without flux is $\phi_0 = 0$ for all plots. The complete density of states is shown for a junction length of (a) $L = 0.5\, \xi_0$, (b) $L = \xi_0$ and (c) $L = 2\, \xi_0$. In (d) the contribution to the density of states of left movers alone is shown.}
	\label{fig:dos}
\end{figure}

\begin{figure}
\centering
	\includegraphics[width=\columnwidth]{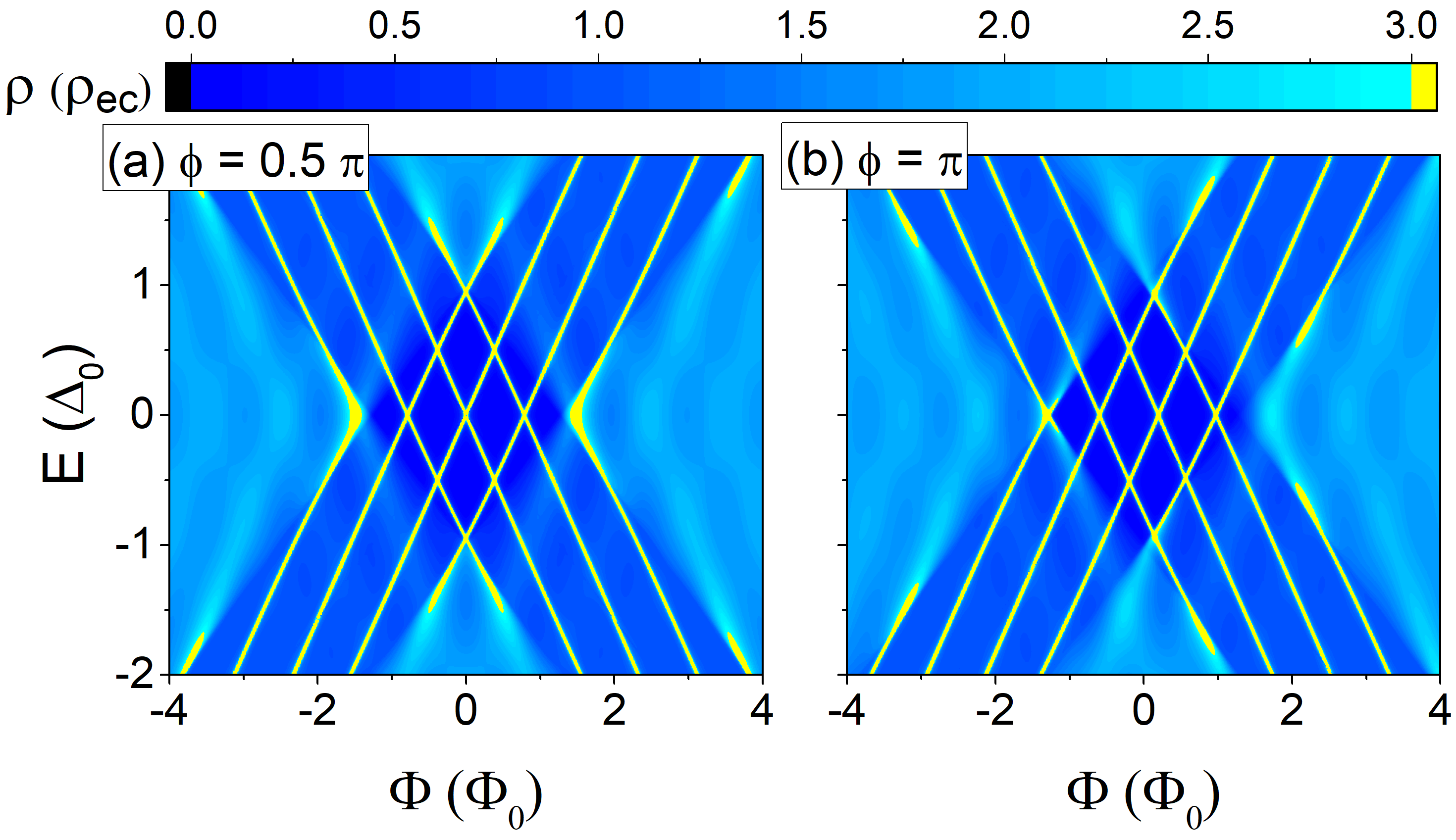}
	\caption{Density of states of the topological Josephson junction for a junction length of $L=2\, \xi_0$ with (a) the superconducting phase difference $\phi_0 = \pi / 2$ and (b) $\phi_0 = \pi$. The $\phi_0$ dependence is $2\,\pi$-periodic and shifts the position of the Andreev bound states.}
	\label{fig:dos_phi}
\end{figure}

Including the contributions for quasiparticles of both types impinging from the right-hand side and performing the sum over momenta, we obtain the density of states $\rho(E)$ of the 2D topological Josephson junction in units of $\rho_\text{ES} = (\pi \hbar v_\text{F})^{-1}$, the density of states per unit volume of a single edge channel. For energies above the superconducting gap, we find
\begin{equation}
\label{eq:dos}
\rho (E) =\sum_{\sigma=\pm}\rho_\text{BCS}(E_\sigma)F_\sigma(E_\sigma)
\end{equation}
where 
\begin{equation}
\rho_\text{BCS}(E) = \frac{|E|}{\sqrt{|E|^2 - \Delta^2}}\Theta(|E|-\Delta),
\end{equation}
is the BCS density of states. The function 
\begin{equation}
\label{eq:interference}
F_\pm(E) = \frac{E^2-\Delta^2}{E^2-\Delta^2 \cos^2(\frac{\phi_u}{2} \pm \frac{E L}{\Delta \xi_0})}
\end{equation}
is a modulating function which arises due to quantum interference. We underline that the energy-dependent term inside the cosine arises from the energy dependence of the electron-like and hole-like wave vectors and, from a physical point of view, reflects an additional phase picked up by an electron-hole pair making a round-trip through the junction. Note that the density of states inside the gap can be obtained from Eq.~\eqref{eq:dos} in a standard way via analytic continuation $E\to E+i0^+$, and can be expressed in an analogous form.

The proximity induced density of states, cf. Fig.~\ref{fig:dos} shows interesting features that are worth discussing in detail. As one can argue from the above expression, the full density of states is given by a sum of two BCS-like contributions, shifted in energy in opposite directions, with a modulating factor closely linked to quantum interference effects.

Looking at Fig.~\ref{fig:dos} one can see that in the central diamond energies are smaller than the superconducting gap and both channels are closed. Inside this regions sharp discrete peaks appear in the density of states due to the formation of Andreev bound states inside the junction.
In the dark blue, diagonal arms, one of the two channels is opened due to an interplay between the particle energy and the Doppler shift that arises from the applied flux. The slope $\alpha$ of the arms is inversely proportional on the device length, $\alpha = \pm \frac{1}{L} \frac{\pi \xi_0 \Delta}{v_\text{F} \Phi_0}$.

For large magnetic fluxes, light blue regions form. Here, both channels are opened as the Doppler shift becomes larger than the gap. In addition, for large energies both channels are open since the electron-like (hole-like) energy exceeds the induced superconducting gap. An additional modulation appears in the density of states (more noticeable for long junctions) which stems from the energy dependence of the electron-like and hole-like wave vectors. The slope of these oscillations is twice the one of the arms themselves.

The density of states depends on the various system parameters in a nontrivial way. Increasing the junction length $L$ while keeping its width $W$ and the magnetic field $B$ fixed leads to a linear increase of $\Phi$. Furthermore, both the frequency and the strength of the oscillations arising from the energy-dependence of wave vectors increase.
Similarly, increasing the junction width $W$ at constant $L$ and $B$ gives rise to an increased flux as well as to a growing Cooper pair momentum $p_\text{S}$. Finally, changing $B$ at fixed $L$ and $W$ yields a change of both the magnetic flux and the Cooper pair momentum. 
Further control over the density of states can be obtained by tuning the phase difference $\phi_0$. It can be exploited to move the BCS singularities in the density of states as shown in Fig.~\ref{fig:dos_phi}. Experimentally, $\phi_0$ can be controlled by imposing a supercurrent between the superconducting leads, or alternatively by closing the two leads in a loop. In the latter case the magnetic field also determines $\phi_0$ through flux quantization. In the following discussion, for sake of clarity, we restrict the discussion to $\phi_0=0$, the extension to finite $\phi_0$ being straightforward.

\section{Device characteristics}
\label{sec:characterization}

\begin{figure*}
\centering
	\includegraphics[width=0.8\textwidth]{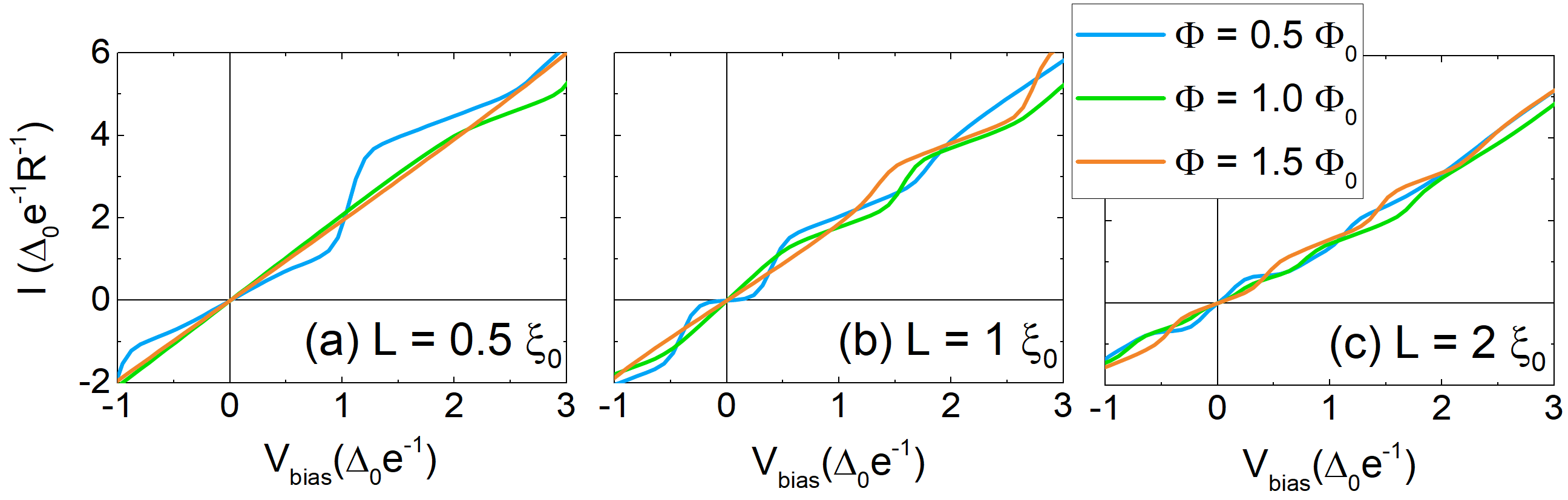}
	\caption{I-V characteristics at various fluxes. Different panels correspond to different junction lengths: (a) $L=0.5\,\xi_0$, (b) $L=\xi_0$ and (c) $L=2\xi_0$. All curves are calculated for a temperature $T / T_C = 0.1$ where $T_C$ denotes the critical temperature of the superconductors. The current and bias voltage are normalized in units of the superconducting gap at zero temperature $\Delta_0$, the electron charge $e$ and the tunnel junction resistance $R$.}
	\label{fig:iv}
\end{figure*}

In order to probe the density of states in the topological edge channels, we consider a normal metal probe coupled to the side of the TI, see Fig.~\ref{fig:device}. We will focus on a weakly tunnel-coupled probe with a constant density of states equal to unity. In the following, we thus characterize the proposed TSQUIPT and its performance in view of future implementation by analysing the charge current injected from the probe terminal for different lengths, temperatures, fluxes and different biasing conditions. 
The general expression for the charge current injected from the tunnel probe is given by~\cite{bardeen_tunnelling_1961}
\begin{equation}
I(V)=\frac{1}{e R}\int dE\; \rho(E)\left[f_\text{N}(E)-f_\text{S}(E)\right]
\end{equation}
where $f_i(E)=\{\exp[(E-\mu_i)/(\kBT)]+1\}^{-1}$ denotes the Fermi functions of the normal probe ($i=\text{N}$) and the superconductors ($i=\text{S}$), respectively, and $R$ is the resistance of the tunnel barrier. 
The resulting charge current is plotted in Fig.~\ref{fig:iv} for various junction lengths and fluxes. Due to the Andreev bound states, the conduction is almost never zero. It is shown that a smaller flux is required to open both channels in the case of short junctions. This is a result of comparing junctions of different length, but of the same width. Thus, the same flux through a shorter junction translates to a higher magnetic field and a higher Cooper pair momentum $p_s$. In the regions where only one channel contributes to transport, the slope $dI/dV$ is halved.

\begin{figure}
\centering
	\includegraphics[width=\columnwidth]{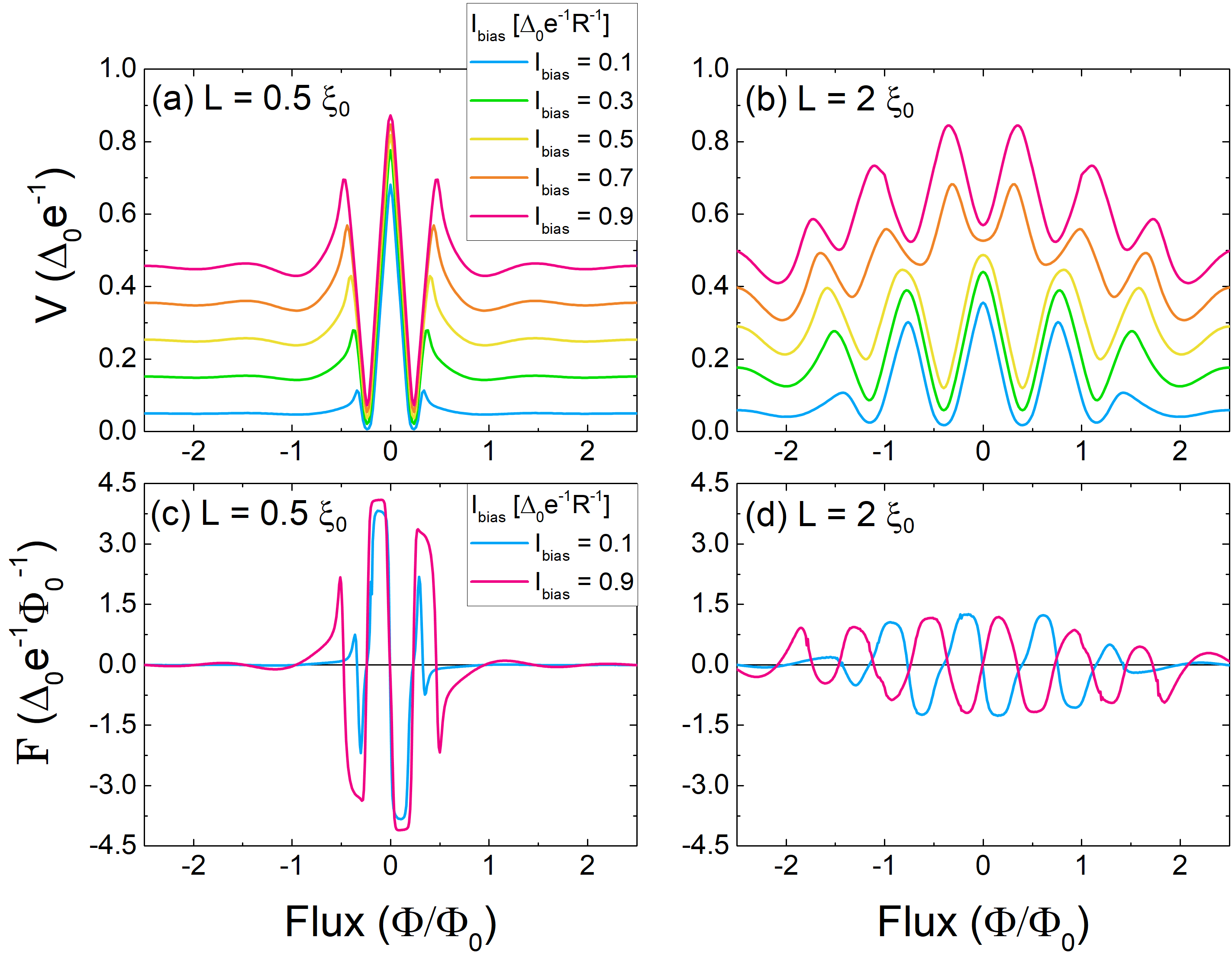}
	\caption{(a) and (b) show the voltage drop across the device as a function of the flux for several values of the bias current, for junctions of length $L=0.5\,\xi_0$ and $L=2\, \xi_0$ respectively. Figures (c) and (d) show the corresponding transfer function $dV/d\Phi$ for high and low bias currents, for junctions of length $L=0.5\,\xi_0$ and $L=2\,\xi_0$ respectively. All curves are calculated for a temperature $T / T_C = 0.1$.}
	\label{fig:vflux_transfer_function}
\end{figure}

in the following, we restriced our selves to the case in which the device is operated via a current bias for the sake for clarity. Figure~\ref{fig:vflux_transfer_function} shows the potential difference between the probe and the superconductors as a function of flux for several bias currents, as well as the voltage to flux transfer function
\begin{equation}
{\cal F} = \frac{\partial V}{\partial \Phi}~,
\end{equation}
for junctions of different length. The calculated transfer function compares favourably to a state of the art SQUIPT~\cite{dambrosio_normal_2015}, which can reach up to $\unit[0.4]{mV}/\Phi_0$ or $2 \frac{\Delta_0}{e \Phi_0}$, with $\Delta_0 \approx \unit[200]{\mu eV}$, the gap of aluminium taken as a reference. The transfer functions change sign multiple times, and are damped at higher fluxes.

\begin{figure}
\centering
	\includegraphics[width=0.8\columnwidth]{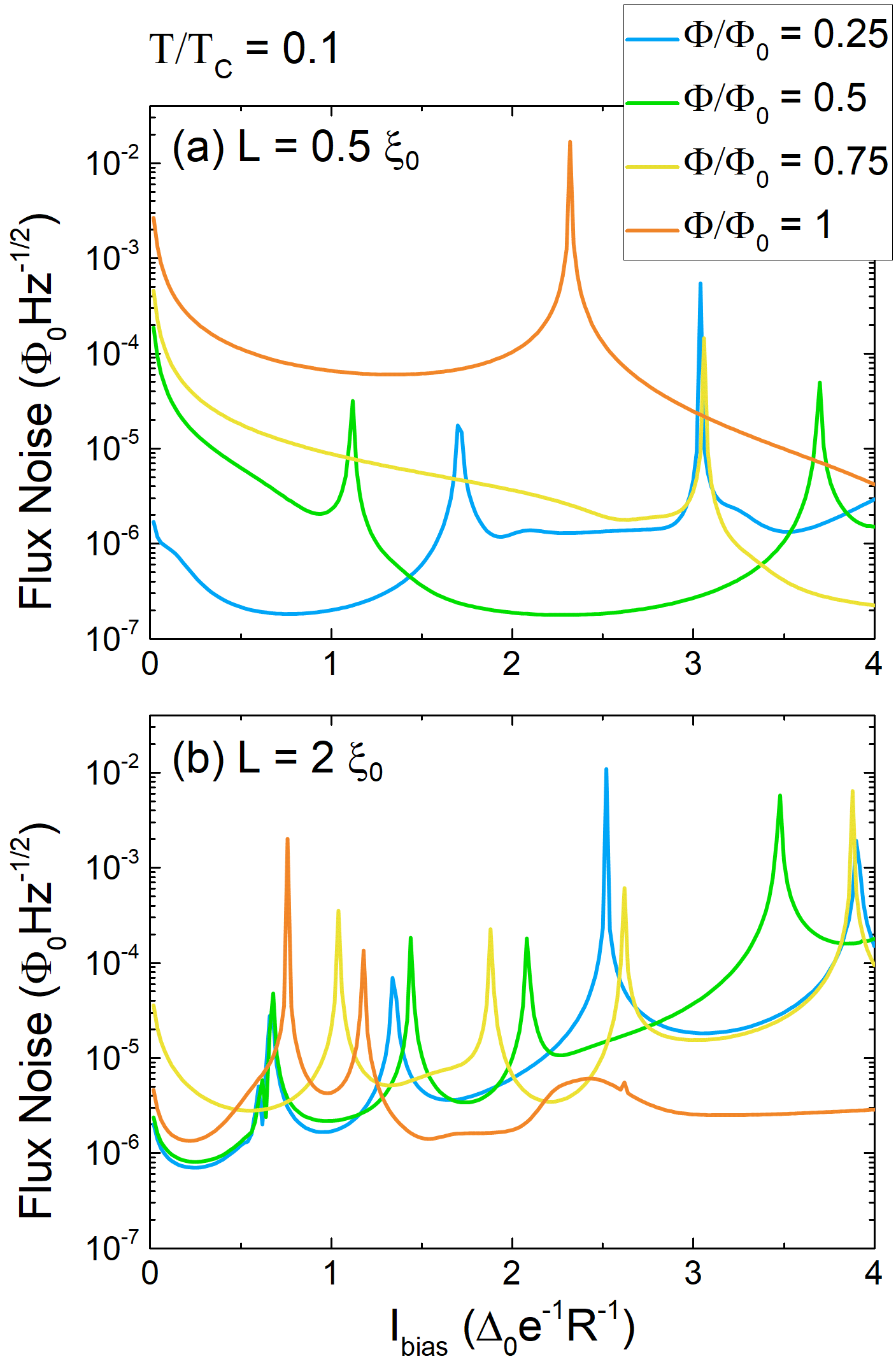}
	\caption{The figure shows the flux noise for various fluxes at a temperature of $T / T_C = 0.1$ and as a function of the bias current. (a) For a junction of length $L=0.5\,\xi_0$ and (b) $L=2\,\xi_0$.}
	\label{fig:flux_noise}
\end{figure}

\begin{figure}
\centering
	\includegraphics[width=0.8\columnwidth]{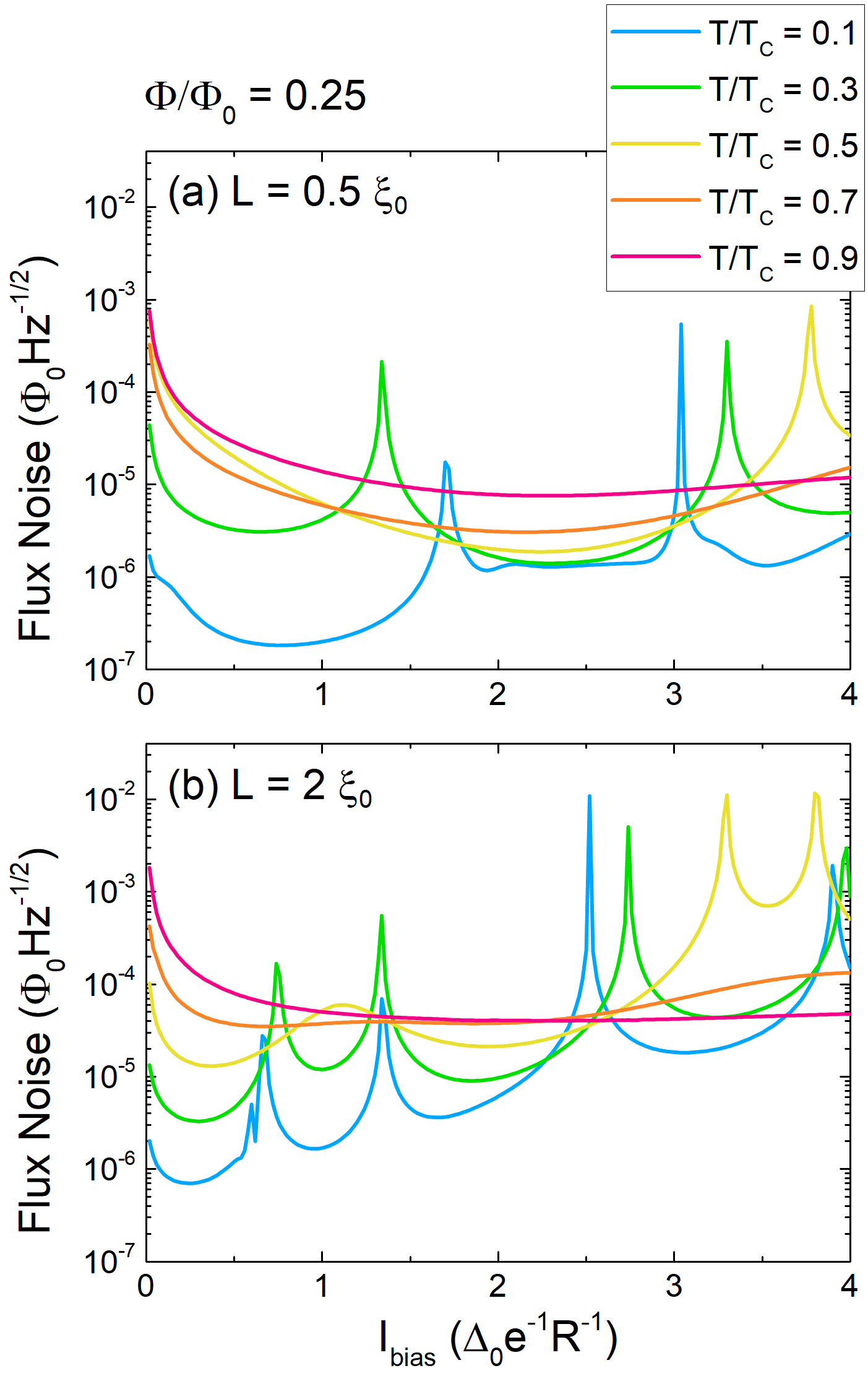}
	\caption{The temperature dependence of the flux noise is shown for a flux of $\Phi / \Phi_0 = 0.25$ as a function of the bias current for (a) a junction of length $L=0.5\,\xi_0$ and (b) $L=2\,\xi_0$.}
	\label{fig:flux_noise_T_dependence}
\end{figure}

As shown in Fig.~\ref{fig:vflux_transfer_function}, the voltage drop across the device depends strongly on the magnetic flux for low flux, and tends to a finite value at high flux. As the flux dependence of the TSQUIPT is not periodic, the proposed 2D topological Josephson junction device, can be utilized as a new type of \emph{absolute} magnetometer. That is, by tracing $V(\Phi)$ function, one can determine the absolute flux, and by extension the magnetic field to which the device is exposed. This is in stark contrast with conventional SQUID-based magnetometers, which have a $2\,\pi$ periodic flux dependence. To quantify the device's flux sensitivity, we calculate the flux noise given by
\begin{equation}
\phi_{ns} = \frac{\sqrt{S_v}}{|{\cal F}(\Phi)]},
\end{equation}
where $S_v = \frac{\partial V}{\partial I} S_I$ and $S_I = 2 e I$ coth$(\frac{eV}{2k_BT})$ is the tunnelling current noise, and ${\cal F}(\Phi)$ is the voltage to flux transfer function. The flux noise is shown in Fig.~\ref{fig:flux_noise} for two junctions of different lengths and for several values of the flux, as a function of the applied bias current. Figure~\ref{fig:flux_noise_T_dependence} shows the temperature dependence of the flux noise. The minimum value of the flux noise is of the order of $10^{-6}$ $\Phi_0 /$Hz$^{1/2}$. The curves exhibit large peaks, which are a consequence of the transfer function ${\cal F}(\Phi)$ regularly crossing zero. Note that the flux noise $\phi_{ns} \propto \sqrt{R}$, the tunnel junction resistance, which we have taken to be 10 k$\Omega$.

One big advantage of the considered device is that, contrary to conventional SQUID and SQUIPT designs, no ring structure is needed for its implementation. The noise depends strongly on the zeroes in the transfer function, which can be controlled through $\phi_0$ via an imposed supercurrent, allowing for an easily optimized performance of the proposed device. The possibility to control the superconducting phase difference via a current bias is another advantage over conventional SQUID designs. While the behaviour of the flux noise depends on the junction length, width, the applied flux and the temperature in a complicated way, the general rule is that the shorter junction perform better. The performance is ultimately limited by the technical ability to fabricate the junctions, and using state-of-the-art techniques the device has the potential to be extremely small, as the fundamental size limit is determined by the topological edge channel width.

\section{Summary and conclusions}
\label{sec:conclusion}
We have proposed a Topological SQUIPT, based on a 2D TI in close proximity to two superconductors, designed to investigate the subtle interplay between superconducting currents that arise in presence of a small magnetic field, and the helical edge states of a 2D TI. A weakly tunnel coupled normal probe allows us to inspect the edge channel density of states which has a non-trivial energy and flux dependence. We conclude that the proposed device can be operated as a new type of sensitive, absolute magnetometer, that features several advantages over conventional SQUID and SQUIPT designs.

Understanding and verifying the electrical characteristics of this device will reinforce our knowledge of structures based on S-TI interfaces, and pave the way for more complex experiments that aim to exploit quantum interference phenomena present in these hybrid structures.

The proposed device is also interesting from a spintronics perspective; when only one channel is open, the current is spin polarized, due to the spin-momentum locking. This effect could be observed by replacing the normal metal probe with a spin selective probe, e.g., a ferromagnetic tunnel coupled probe.

\begin{acknowledgments}
We acknowledge financial support from the Ministry of Innovation NRW. E. Strambini and F. Giazotto acknowledge financial support from ERC grant agreement no. 615187 – COMANCHE. 
The work of F. G. was partially funded by the Tuscany Region under the FARFAS 2014 project SCIADRO. M. C. acknowledges support from the CNR-CONICET cooperation programme ``Energy conversion in quantum, nanoscale, hybrid devices''. E.M.H. and L.W.M.   acknowledge   financial   support   from the German Science Foundation (Leibniz Program,  SFB1170  "ToCoTronics")  and  the  Elitenetzwerk  Bayern  program  ``Topologische Isolatoren''. 

\end{acknowledgments}

% \bibliographystyle{apsrev4-1}
% \bibliography{/home/bjoern/LaTeX/Bibtex/Meine_Bibliothek}


\end{document}